\documentclass[aps,prb,twocolumn,showpacs,psfig,superscriptaddress]{revtex4-1}
\usepackage{times}
\usepackage{graphicx}
\usepackage{float}
\usepackage{latexsym,amsmath,amssymb,bm,euscript}
\usepackage{color}
\usepackage{subfigure}
\usepackage{epstopdf}
\usepackage[colorlinks=true,linkcolor=blue,citecolor=blue]{hyperref}
\usepackage{hyperref}
\usepackage{soul}
\usepackage{ulem}
\usepackage{mathrsfs}
\usepackage{amsmath}
\usepackage{CJK}

\begin{document}
\begin{CJK*}{GBK}{song}
\title{Bilayer Linearized Tensor Renormalization Group Approach for Thermal Tensor Networks}

\author{Yong-Liang Dong}
\affiliation{Department of Physics, Key Laboratory of Micro-Nano Measurement-Manipulation and Physics (Ministry of Education), Beihang University, Beijing 100191, China}

\author{Lei Chen}
\affiliation{Department of Physics, Key Laboratory of Micro-Nano Measurement-Manipulation and Physics (Ministry of Education), Beihang University, Beijing 100191, China}

\author{Yun-Jing Liu}
\affiliation{Department of Physics, Key Laboratory of Micro-Nano Measurement-Manipulation and Physics (Ministry of Education), Beihang University, Beijing 100191, China}

\author{Wei Li}
\email{w.li@buaa.edu.cn}
\affiliation{Department of Physics, Key Laboratory of Micro-Nano Measurement-Manipulation and Physics (Ministry of Education), Beihang University, Beijing 100191, China}
\affiliation{International Research Institute of Multidisciplinary Science, Beihang University, Beijing 100191, China}

\begin{abstract}
In this paper, we perform a comprehensive study of the renormalization group (RG) method on thermal tensor networks (TTN). By Trotter-Suzuki decomposition, one obtains the 1+1D TTN representing the partition function of 1D quantum lattice models, and then employs efficient RG contractions to obtain the thermodynamic properties with high precision. The linearized tensor renormalization group (LTRG) method, which can be used to contract TTN in an efficient and accurate way, is briefly reviewed. In addition, the single-layer LTRG can be generalized to a bilayer form, dubbed as LTRG++, in both finite- and infinite-size systems, with accuracies significantly improved. We provide the details of LTRG++ in finite-size system, comparing its accuracy with single-layer algorithm, and elaborate the infinite-size LTRG++ in the context of fermion chain model. We show that the LTRG++ algorithm in infinite-size system is in essence equivalent to transfer-matrix renormalization group method, while expressed in a tensor network language. LTRG++ is then applied to study an extended Hubbard model, where the phase separation phenomenon, groundstate phase diagram, as well as quantum criticality-enhanced magnetocaloric effects under external fields, are investigated.

\end{abstract}
% 02.70.-c  Computational Techniques
% 05.30.-d  Quantum Statistical Mechanics
% 05.10.Cc  Renormalization group methods
% 75.10.Jm  Quantized spin models, including quantum spin frustration
\maketitle
\end{CJK*}

\section{Introduction}
Renormalization group methods constitute an important numerical approach to calculate quantum lattice models. The density matrix renormalization group (DMRG) method \cite{DMRG, DMRG-Review} and its developments in the age of tensor networks (TNs) \cite{DMRG-MPS} have become the method of choice while calculating ground (or low-energy excited) state properties of 1D and quasi-1D lattice models. In the TN language, the ground state $| \Psi_g \rangle$ can be expressed as a TN state (matrix-product state in 1D and tensor-product state in 2D or higher), which can then be optimized either according to energy criterion, i.e., minimal energy for ground state, or by a imaginary-time evolution which evolves from a random initial state gradually to the desired ground state of a given manybody Hamiltonian, in 1D chains,\cite{iTEBD} 2D lattices,\cite{Jiang-2008} or infinite-dimensional Bethe lattices.\cite{Li-2012}

Apart from ground state properties, we are also interested in computing thermodynamic properties of quantum manybody systems, especially when a comparison or fitting between model simulation and experimental measurement is required. Since any experimental measurement is performed at finite temperatures, a thermal TN (TTN) simulation is thus indispensable. In fact, when the temperature is sufficiently low, one can approximate well the ground state properties with low-temperature results (see, for instance, Sec. \ref{Sec:HubbardGround} below). More interestingly, some exotic thermal effects at low temperature, strongly affected by quantum fluctuations, can also be investigated by TTN calculations. For instance, a quantum critical point (QCP) will spread through a finite-temperature ``critical" region, where the thermal and dynamical properties are strongly influenced by the QCP.\cite{Sachdev} Although the quantum phase transition is present only at absolutely zero temperature, and even does not lead to any ``true" phase transition at finite temperature, it could lead to some nontrivial consequences in this exotic region. To be more concrete, among others, the entropy in response to external fields will show a singular behavior at critical point $B_c$, resulting in a QCP-enhanced magnetocaloric effect (MCE).\cite{Zhu-2003,Garst-2005,Rost-2009,Ryll-2014,Sharples-2014} In the vicinity of QCP, the magnetic Gr\"uneisen parameter $\Gamma_B = (\frac{\partial T}{ \partial B})_S$ measuring the MCE, shows a diverging peak as temperature goes to zero, whose scaling behavior is intimately related to the quantum criticality.\cite{Zhu-2003}

\begin{figure}[tbp]
  \includegraphics[angle=0,width=1\linewidth]{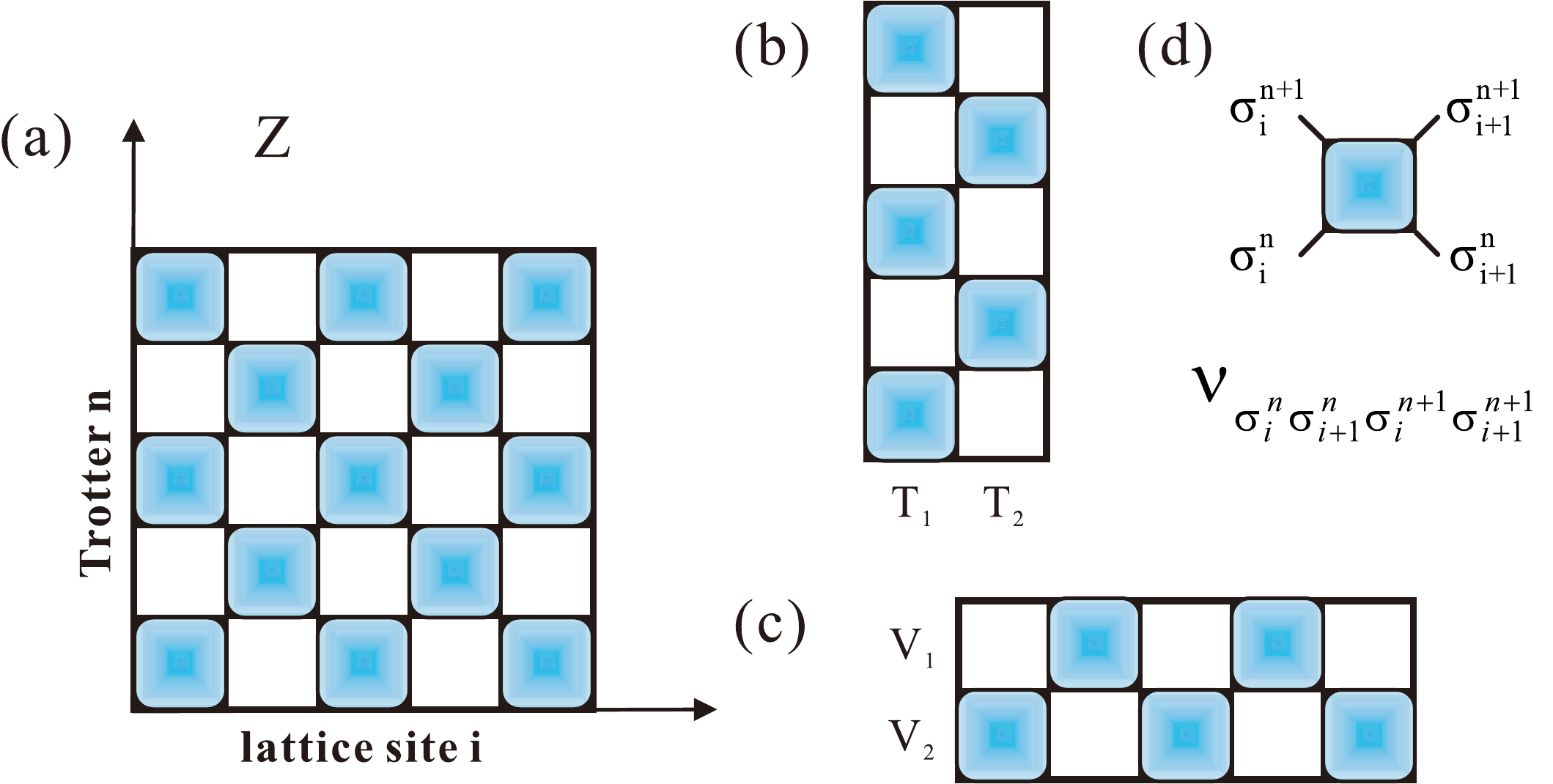}
  \caption{(\textbf{a}) The 1+1D TTN represents the partition function $\bold{Z}$ of a 1D quantum lattice model, which exhibits a checkerboard pattern. (\textbf{b}) and (\textbf{c}) are the transfer matrices along spatial and Trotter directions, respectively. (\textbf{d}) depicts the rank-four local tensor $\nu$, the elementary unit in the TTN.}
  \label{Fig:TTN}
\end{figure}

In order to calculate the thermal properties with RG algorithms, one turns to density matrix $\rho = e^{-\beta H}$ describing a mixed state, instead of a pure state $| \Psi \rangle$. Interestingly, the density matrix $\rho$ also bears a TN representation, and can be treated with TN-based algorithms. This TTN can be obtained by expanding $\rho$ with Trotter-Suzuki decomposition, illustrated as an checkerboard lattice in Fig. \ref{Fig:TTN}(a). For infinite-size systems, Xiang and Wang \cite{TMRG} applied White's DMRG to the finite-temperature systems and developed the transfer-matrix renormalization group (TMRG) method. TMRG can determine the thermodynamic properties with very high accuracy, and has been widely accepted as the standard method to calculate thermodynamics of strongly correlated quantum lattice systems in 1D. \cite{TMRGApp}

Motivated by the idea of tensor network, Li \textit{et al.} proposed a linearized tensor renormalization group (LTRG) method,\cite{LTRG} which project the transfer matrix $V_{1(2)}$ continually [see Fig. \ref{Fig:TTN}(c)] to the density matrix of the system (in a matrix product operator form), and cool down the system to various finite temperatures. Considering that the two directions of thermal TN are inequivalent, which is that the spatial direction is infinite while the thermometric axis is finite and subject to a periodic boundary condition, a quite natural contraction procedure is to contract the TTN linearly along the Trotter direction. LTRG method can be used to accurately calculate the thermodynamics in 1D chains \cite{Yan-2012} and also in higher dimensional lattices.\cite{Ran-2012}

In this paper, we discuss a variant of the LTRG method, i.e., bilayer LTRG (for short, LTRG++). LTRG++ can achieve a better accuracy than previous single-layer LTRG algorithm. %which is found to %have almost the same accuracy as that of TMRG. Furthermore, we reveal that this is not a coincidence:
Furthermore, although being very different in the arithmetic level, LTRG++ is in essence equivalent to TMRG method, except that it is written in TN language and more straightforward to be implemented in practice. We apply LTRG++ algorithm to study the finite-size Heisenberg systems by slightly adapting the algorithm, as well as the extended Hubbard chain with attractive inter-site coupling by taking fermionic statistics into account.

The rest part of the paper is organized as followings: In Sec. \ref{Sec:TTN}, we will briefly review the Trotter-Suzuki TTN and its LTRG contraction algorithm. The bilayer form of LTRG, adapted to simulate finite-size system, is discussed in Sec. \ref{Sec:FiniSize}. In Sec. \ref{Sec:LTRG++} we introduce the LTRG++ algorithm for infinite chain and show that there exists an intimate relation between LTRG++ and TMRG algorithms. Lastly, in Sec. \ref{Sec:Hubbard} we elaborate some adjustments in LTRG++ to take care of fermionic sign, and then apply it to investigate a fermion extended Hubbard chain model, where the ground state phase diagram, finite-temperature entropy change under magnetic fields, and QCP-enhanced MCE at low temperatures, etc., are explored.

\section{Trotter-Suzuki Thermal Tensor Network and its Linearized Renormalzation Group}
\label{Sec:TTN}

From the tensor-network point of view, the success to calculate the thermodynamics of the system can be ascribed to an efficient contraction of the 1+1D TTN shown in Fig. \ref{Fig:TTN}(a). The 1+1D TTN for $L$-site quantum chain models with Hamiltonian $H = \sum_{\langle i, i+1 \rangle} h_i$ can be obtained by employing the Trotter-Suzuki decomposition.\cite{Suzuki-1976} Take the spin-1/2 Heisenberg system, i.e., $h_i = \vec{S}_i \cdot \vec{S}_{i+1}$, as an example, $\rho = [\exp{(-\tau H)}]^N$, where $N \tau = \beta$. Via the (first-order) Trotter-Suzuki decomposition, it can be further expressed as
\begin{equation}
\rho \approx \sum_{\{\sigma_i^n\}} \prod_{i=1}^{L} \prod_{n=1}^{N} \nu_{\sigma_i^n, \sigma_{i+1}^n, \sigma_i^{n+1}, \sigma_{i+1}^{n+1}},
\end{equation}
by inserting $N$ sets of orthonormal basis $\{ \sigma_i^n \}$ where $i$ is spatial index and $n$ the imaginary-time (Trotter) index. The rank-four tensors can be expressed as follows:
\begin{equation}
\nu_{\sigma_i^n, \sigma_{i+1}^n, \sigma_i^{n+1}, \sigma_{i+1}^{n+1}} = \langle \sigma_i^{n}, \sigma_{i+1}^{n} | e^{-\tau h_i} | \sigma_{i}^{n+1}, \sigma_{i+1}^{n+1}\rangle
\end{equation}

In order to calculate the thermodynamic properties, one needs to contract this TTN accurately, which, unfortunately, is an NP-hard problem and thus can not be solved exactly. Therefore, people has to resort to methods to approximate efficient contractions of TTN. Among others, RG-based algorithms, including TMRG and LTRG for infinite chains, finite-temperature DMRG\cite{Feiguin-2005}, and minimally entangled typical thermal state (METTS)\cite{METTS} for finite-size systems, etc., constitute an important class of approaches.

Firstly, we review the TMRG algorithm in brief. As shown in Fig. \ref{Fig:TTN}(b), the TTN can be decomposed into repeated 1D stripes, where the vertical stripe, denoted as transfer matrices $T_1$ and $T_2$, transfers the states ${\sigma_i^n}$'s between different lattice sites. Since $Z = \lim_{N \to \infty} \rm{Tr} (T^{N/2})$, the transfer matrices $T = T_1 T_2$ are of great interest to us, whose dominant eigenvalue in the thermodynamic limit $\lambda_{\rm{max}}$ determines the free energy ($F = \frac{1}{2\beta} \ln \lambda_{\rm{max}}$) and thus other thermodynamic quantities of the 1D quantum system. In order to find this dominant eigenvalue (and corresponding eigenvector), TMRG algorithm incorporates the DMRG method \cite{DMRG} to solve the dominant eigenvalue problem in transfer matrix $T$. The key idea of TMRG is to regard the transfer matrix $T$ as the ``Hamiltonian" in ground state problem, perform ``real space" renormalization along the Trotter direction, and truncate the accumulated $\{\sigma_i\}$ indices into a fixed dimension.

\begin{figure}[tbp]
  \includegraphics[angle=0,width=1\linewidth]{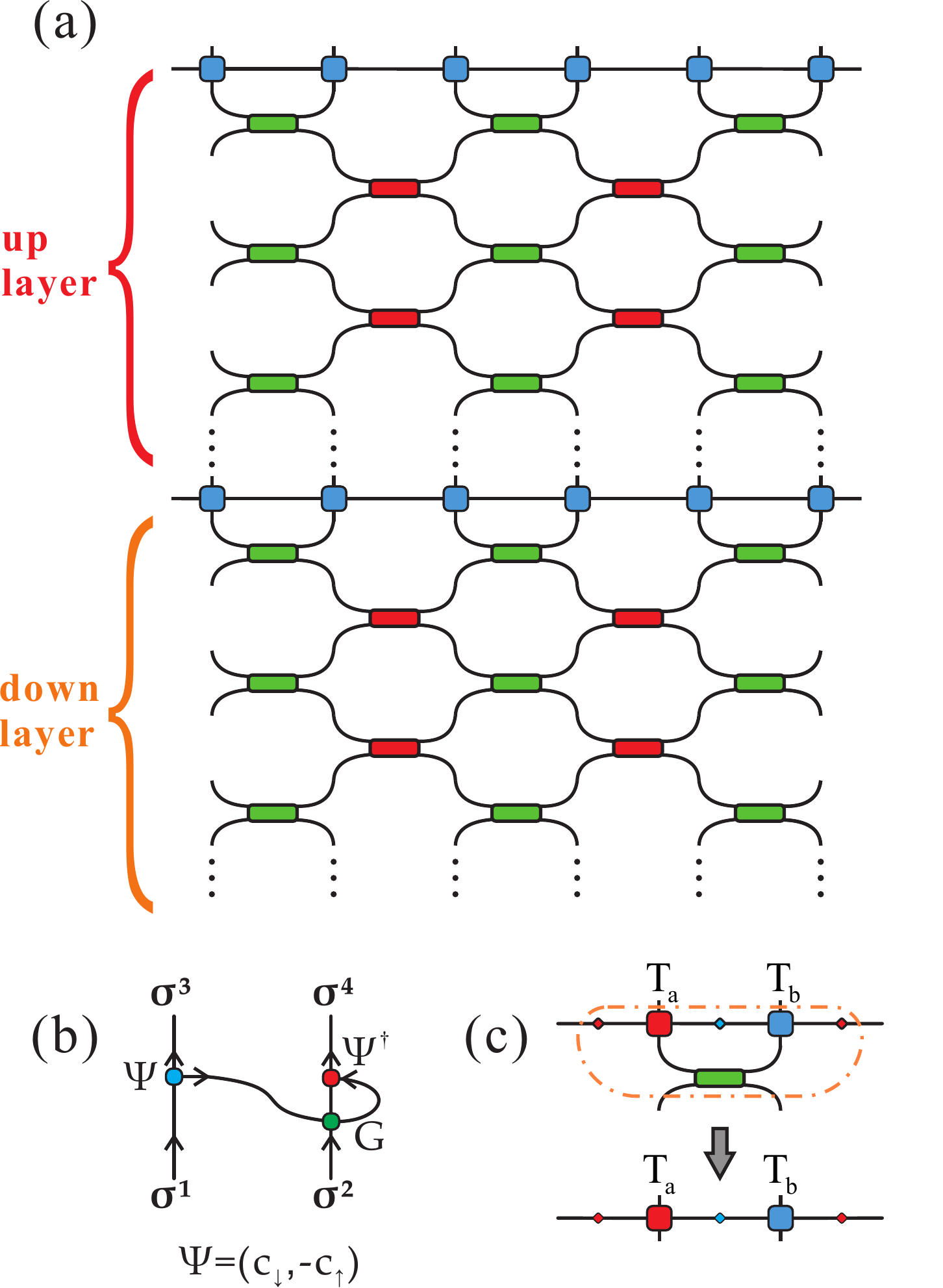}
  \caption{(a) LTRG++ method adopts a symmetric construction, i.e., the upper and lower layers are identical. (b) A local hopping term $\Psi_i \cdot \Psi^{\dagger}_j$, where $\Psi = (c_{\uparrow}, c_{-\downarrow})$ constitutes an irreducible representation of $SU(2)_{spin}$ symmetry. Note that there is a matrix $G$ at the cross point of the Jordan-Wignar string and the incoming physical index, which takes care of the exchange fermion sign. (c) Local projection and truncation scheme.}
  \label{Fig:LTRG++}
\end{figure}

\begin{figure}[tbp]
  \includegraphics[angle=0,width=1\linewidth]{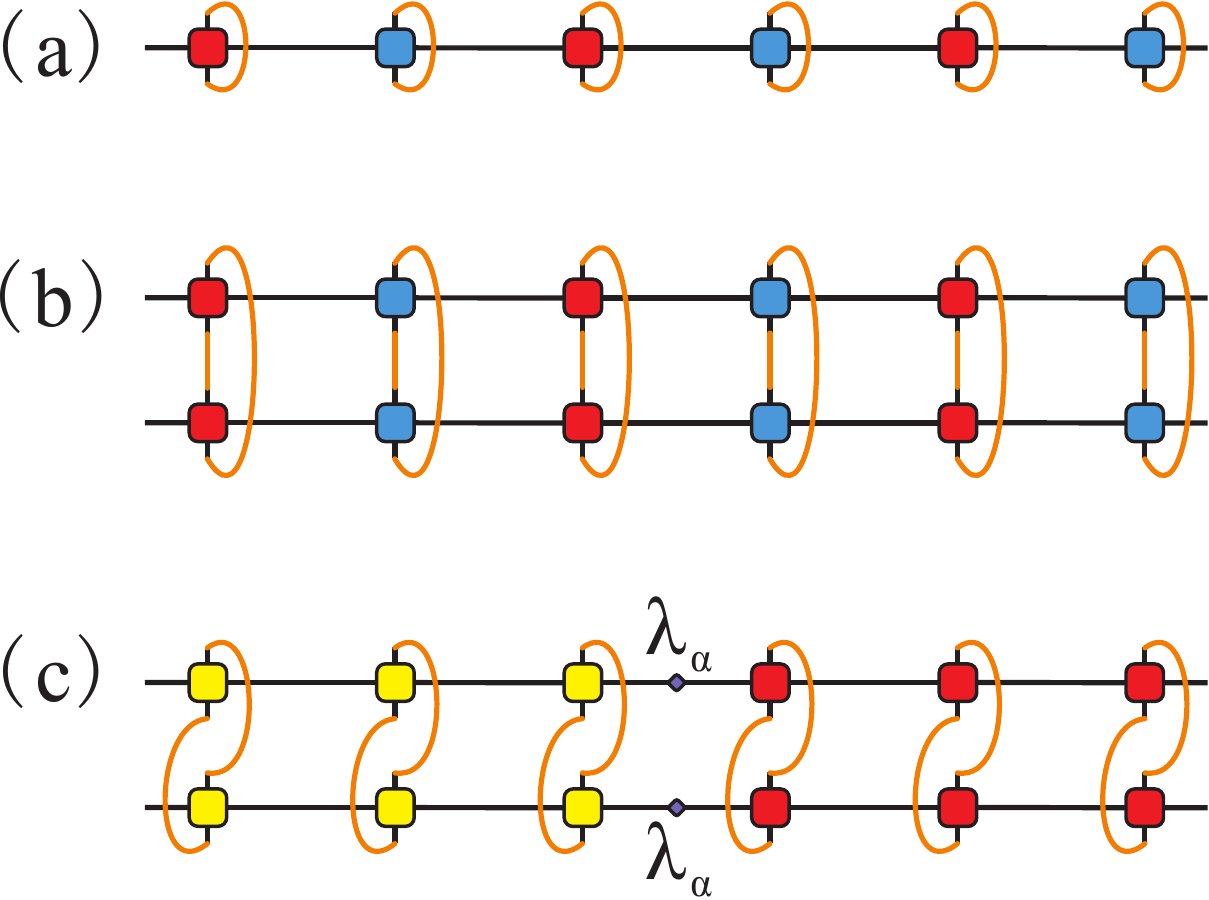}
  \caption{(a) Tensor trace of a single MPO, (b,c) two different ways to contract MPO with its conjugate.}
  \label{Fig:TrMPO}
\end{figure}

Alternatively, efficient contractions of the 1+1D TTN can be also performed by making use of the horizontal transfer matrices $V_1, V_2$ [see Fig. \ref{Fig:TTN} (c)]. In the LTRG algorithm, only a single MPO [upper or lower one in Fig. \ref{Fig:LTRG++}(a)] is involved in the process of contractions: Starting from an identity MPO which represents the density matrix at infinite high temperature ($\beta=0$), and by continually projecting $\nu$ tensors onto the MPO, we cool down the system from $\beta=0$ to various lower temperatures, $\beta=n \tau$ at $n$-th step, and renormalize $\rho$ with factor $\kappa_n$ to regulate the elements with the largest norm in each local tensor (of MPO) as 1. The thermodynamic properties can be computed at temperature $1/\beta$ by collecting these renormalization factors ($\kappa_n$) and by tracing the density matrix MPO ($\rho$). In particular, the partition function is
\begin{equation}
Z(\beta) = (\prod_{n=1}^N \kappa_n^L) \cdot \rm{Tr(\rho)},
\end{equation}
where $\beta=N \tau$ is the inverse temperature, $L$ is the total length of the chain, $\rm{Tr} (\rho)$ is the tensor trace of MPO [see Fig. \ref{Fig:TrMPO}(a)].

An important technique in performing the contractions is to compress the exponentially proliferating bond dimensions of the MPO efficiently. During the process of projecting evolution operators $\nu$ to MPO's, one regards the MPO as a super vector and follows the decimation technique developed in Ref. \onlinecite{iTEBD}. As shown in Fig. \ref{Fig:LTRG++}(c), one contracts $T_a$, $T_b$, projector $\nu$, as well as three diagonal matrices $\lambda_{a,b}$'s, into a single matrix $M$. One then take a singular value decomposition $M = U S V^{\dagger}$, perform truncations according to the singular values in $S$, and reshape the obtained isometries $U,V^{\dagger}$ (multiplied with $1/\lambda_b$) to update the rank-three tensors $\tilde{T}_{a,b}$. Iterating the above procedure, LTRG can cool down the system and produce quite accurate results of thermodynamic quantities even at quite low temperatures, with comparable precision to TMRG.\cite{LTRG}

\section{Finite-Size single- and double-layer LTRG algorithm}
\label{Sec:FiniSize}
In this section, we adapt the LTRG method, which was originally proposed for the infinite-chain system, to a finite-size system, and consider both single-layer and bilayer algorithms. In essence, the finite-size LTRG algorithm is a tensor-network description of the finite-temperature DMRG.\cite{Feiguin-2005} Similarly as in the infinite-chain LTRG, we start with preparing a thermal state at infinitely high temperature, which can be represented by MPO consisting of identity tensors and satisfies the canonical form. We can sequentially project tensor $\nu$ on it and generate the desire finite temperature state expressed in an MPO form, which can be traced to calculate the partition function, as well as other thermal properties.

We would like to emphasize that the single-MPO finite-size LTRG algorithm is different from the finite-temperature DMRG introduced in Ref. \onlinecite{Feiguin-2005}, in which a purified state in an enlarged Hilbert space and its conjugate are involved. In the TTN language, this corresponds to a bilayer LTRG algorithm [see Fig. \ref{Fig:LTRG++}(a)], i.e., we need to contract MPO with its conjugate together to obtain the partition function. It's interesting to find that the bilayer algorithm can produce more accurate result without further computational costs, and more astonishingly, even less than the single-layer one, considering the contraction of MPO with its conjugate is costly [O($\chi^3$)]. The reason is that the canonical form of MPO actually saves us from actually doing this two-MPO contraction. As long as one keeps the canonical form of MPO during the process of projections, the tensor trace of MPO with its conjugate [Fig. \ref{Fig:TrMPO}(c)] is just tracing the square of diagonal matrix $\lambda$, which is costless. Another point is that, by utilizing a bilayer construction, one can actually reduce the number of projection steps by a half.

\begin{figure}
  \includegraphics[angle=0,width=0.85\linewidth]{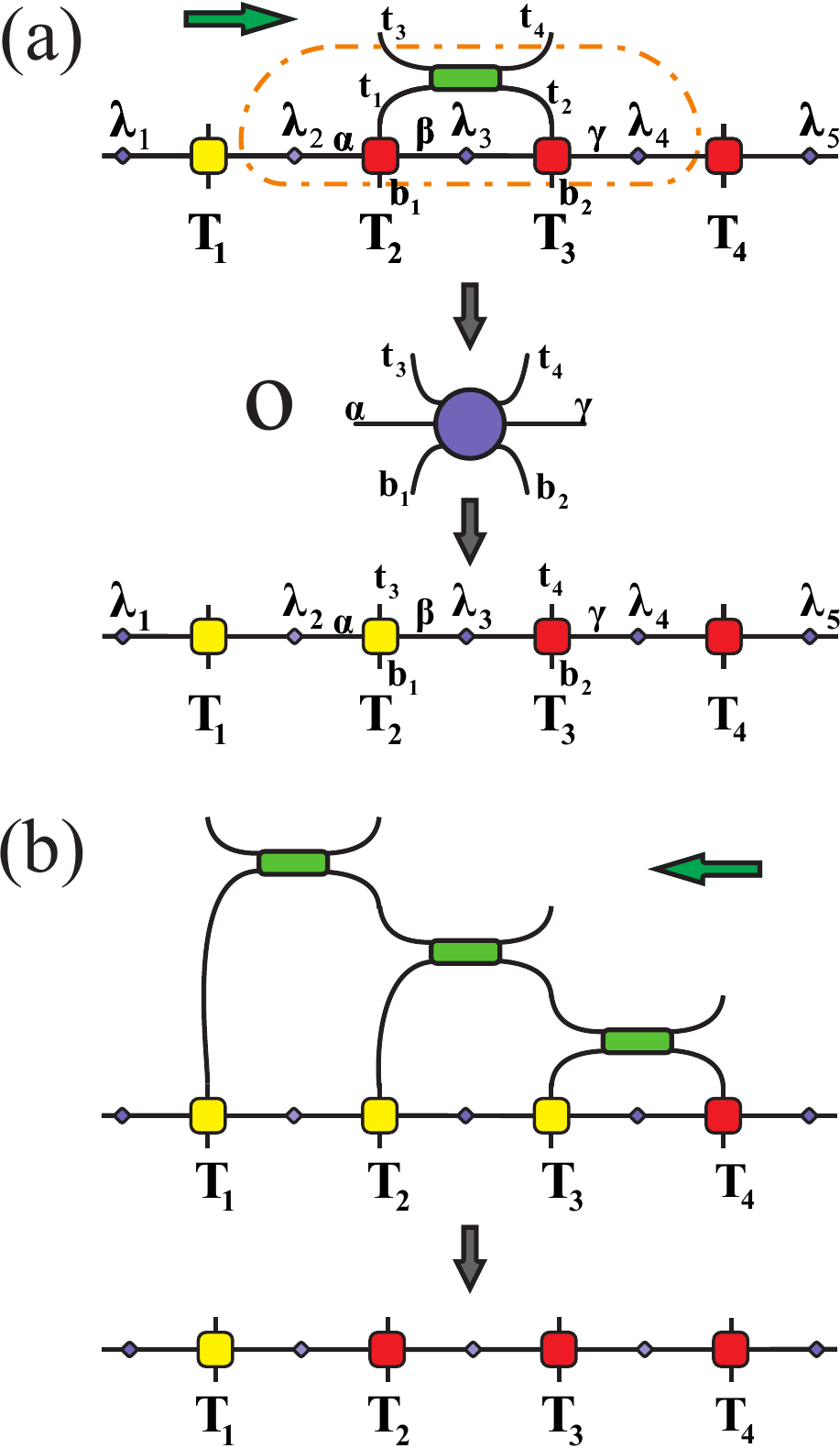}
  \caption{\ (a) Forward sweep from left to right, and the local contraction and decimation scheme. (b) Backward sweep from right to left. Tensors colored yellow(red) satisfy left(right) canonical condition.}
  \label{Fig:FSCanon}
\end{figure}

In the course of projections, we take forward and backward sweeps iteratively and can thus maintain the canonical form, at the same time we compress the bond to avoid exponential proliferation by following details (Fig. \ref{Fig:FSCanon}). In canonical form, each bond is associated with a Schmidt vector $\lambda$ and each local site with a local tensor $T_i$, which satisfies some peculiar relations which will be clarified shortly. As for our projection, take a single step in forward sweep (left to right) on a four-site MPO as an example [Fig. \ref{Fig:FSCanon} (a)], when we do the projection on second bond, we contract the Schmidt vectors $\lambda_2$,$\lambda_3$,$\lambda_4$ with the local tensors $T_2$,$T_3$, as well as the projector $\nu_2$, to produce a sixth-order tensor $O$. Notice that the $\alpha$,$\beta$,$\gamma$ are geometric indices and $t_i$'s, $b_i$'s are physical indices.
\begin{equation}
\begin{split}
O_{\alpha,\gamma,b_1,b_2,t_3,t_4}=\sum_{\beta,\gamma,t_1,t_2}(\lambda_2)_{\alpha}(T_2)_{\lambda,\beta,t_1,b_1}
(\lambda_3)_{\beta}\\
\times(T_3)_{\beta,\gamma,t_2,b_2}(\lambda_3)_{\gamma}\nu_{t_1,t_2,t_3,t_4}.
\end{split}
\end{equation}
After contraction, we reshape O-tensor into a matric form and perform SVD on it, $O_{\alpha b_1 t_3,\gamma b_2 t_4}\simeq\sum_{\beta}^{D_c}U_{\alpha b_1 t_3,\beta}{\lambda'_{\beta}}V_{\beta,\gamma b_2 t_4}^\dagger$, where only the dominant $D_c$ singular values stored in $\lambda'_{\beta}$ are kept. Then we update our local tensor $T_2= (\lambda_2)^{-1} \cdot U$, $T_3=V \cdot (\lambda_44)^{-1}$ and Schmidt vector $\lambda_3=\lambda'$.

It is important to notice that the updated $T_2$ and $T_3$ satisfy left canonical form
\begin{equation}
\sum_{\alpha, b_1, t_3} (\lambda_2 T_2)_{\alpha,b_2,t_3, \beta}  (\lambda_2 T_2)_{\alpha,b_2,t_3, \beta'} = \delta_{\beta, \beta'},
\label{Eq:LeftCanon}
\end{equation}
and right canonical form
\begin{equation}
\sum_{b_2, t_4, \gamma} (T_3 \lambda_3)_{\beta, b_2,t_4,\gamma}  (T_3 \lambda_3)_{\beta',b_2,t_4, \gamma} = \delta_{\beta, \beta'},
\label{Eq:RightCanon}
\end{equation}
respectively. After each projection, we need to normalize the weights in $\lambda_i$ to avoid divergence. As shown in Fig. \ref{Fig:FSCanon}(a), $T_2$ ($T_3$) satisfying left(right) canonical form is represented by yellow (red) tensor.

We proceed to project tensors $\nu$ one after another to cool down the system. Since our initial MPO representing infinite temperature state is canonical, performing the projections from left to right, we can accomplish one full Trotter step, with the canonical form of MPO maintained. Afterwards, as the backward projection proceeds shown in Fig. \ref{Fig:FSCanon}(b), the projection are performed site by site, from right to left, and local tensors will then be changed from ``yellow" to ``red", i.e., from left to right canonical.

\begin{figure}
  \includegraphics[angle=0,width=1\linewidth]{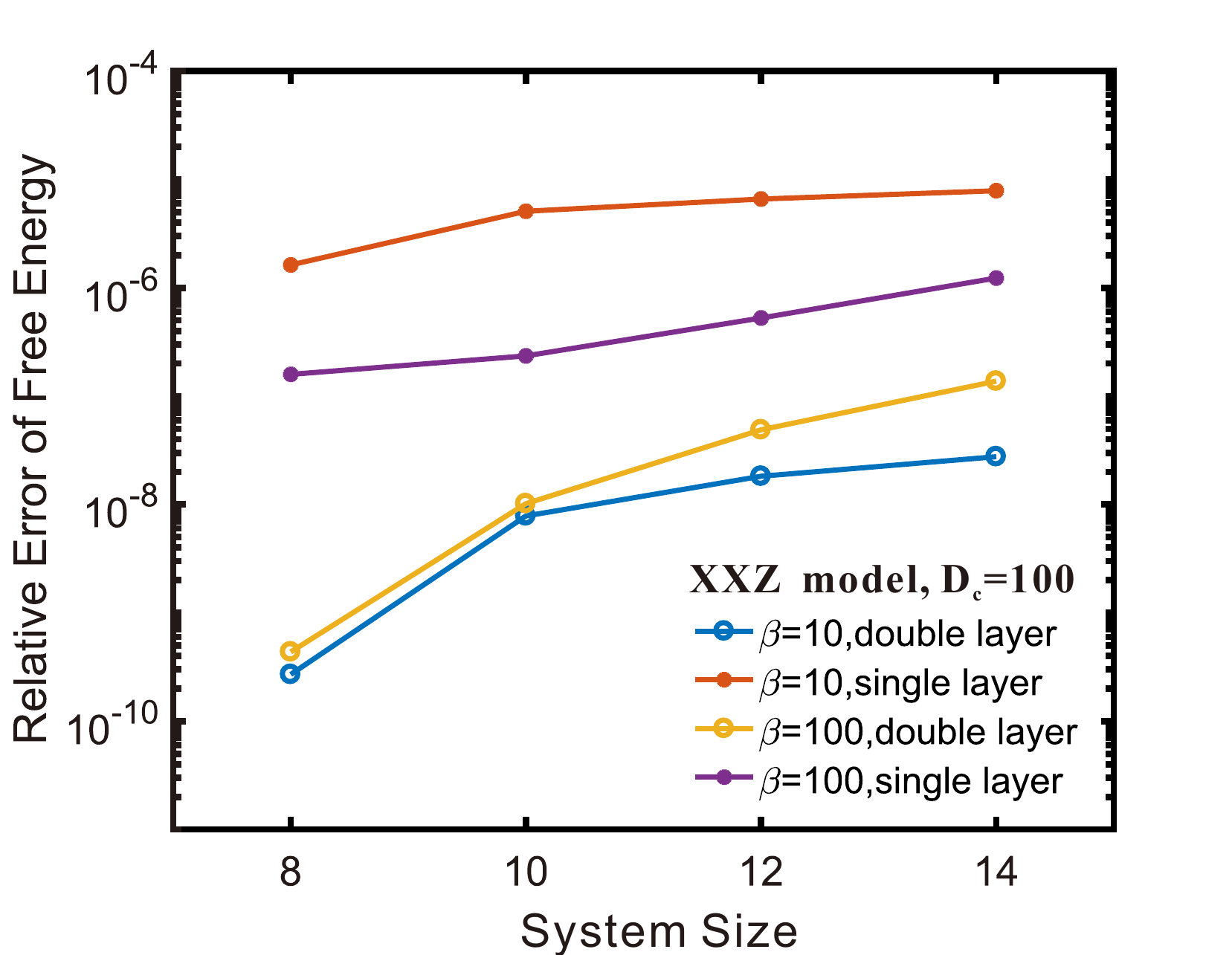}
  \caption{Comparisons of accuracies in single- and double-layer calculations. The vertical axis represents the relative error of free energy to exact diagonalize data.}
  \label{Fig:FSCompare}
\end{figure}

In both the single- and double-layer algorithms, the projection procedure are very similar, while they only differ in the contraction part. In single-layer method, the MPO $e^{-\beta H}$ is prepared and directly traced as illustrated in Fig. \ref{Fig:TrMPO}(a) while in the bilayer approach we prepare MPO $\rho(\frac{\beta}{2}) = e^{-\frac{\beta}{2} H}$ and its conjugate, and take the trace shown in Figs. \ref{Fig:TrMPO}(b,c). Notice that there exists a subtle difference between the contraction schemes in Figs. \ref{Fig:TrMPO}(b) and (c), the former represents $Z = \rm{Tr} [\rho(\frac{\beta}{2}) \cdot \rho(\frac{\beta}{2})]$ and the latter follows $\rm{Tr} [\rho(\frac{\beta}{2}) \cdot \rho(\frac{\beta}{2})^{\dagger}]$. In practical simulations, both schemes provide only marginally different results at high temperatures, while the data are almost the same at intermediate and low temperature region. Nevertheless, the way in Fig. \ref{Fig:TrMPO}(c) is preferred since the canonical condition can be utilized there to avoid us from actually performing the two-MPO contractions. This is explicitly demonstrated in Fig. \ref{Fig:TrMPO}(c), where the two-MPO trace amounts to a trivial sum $\sum_{\alpha} \lambda_{\alpha}^2$, thanks to the favorable environment matrix (i.e. $\mathbb{I}$) to both directions of the bond.

It is thus interesting to compare the accuracy of single- and double-layer algorithms. We take the Heisenberg model as an example, and show the results in Fig. \ref{Fig:FSCompare}. From the plot, we see that at low temperatures (say, $\beta=100$) the bilayer algorithm can achieve two orders of magnitude better accuracy than single-layer method, for various system length $L$. The accuracy of our calculation is significantly boosted of a easy tensor representation by fourth-order Trotter decomposition, which reads:
\begin{equation}
e^{-i\hat H\tau} = \hat U(\tau_1)\hat U(\tau_2)\hat U(\tau_3)\hat U(\tau_2)\hat U(\tau_1), \notag \\
\end{equation}
where $\hat U(\tau_i) = e^{-i\hat H_{\rm{odd}}\tau_i/2}e^{-i\hat H_{\rm{even}}\tau_i}e^{-i\hat H_{\rm{odd}}\tau_i/2}$ and $\tau_1 = \tau_2 = \frac{1}{4 - 4^{1/3}}\tau, \ \tau_3 = \tau - 2\tau_1 - 2\tau_2$.

\section{LTRG++ Algorithm and its Equivalence to TMRG}
\label{Sec:LTRG++}
In Section \ref{Sec:FiniSize}, we showed that the bilayer algorithm in finite-size calculations actually produces a better accuracy than the single-layer algorithm. It is therefore interesting to develop a bilayer algorithm also for infinite-chain system, which will be introduced below in the context of interacting fermions.

\begin{figure}[tbp]
  \includegraphics[angle=0,width=1\linewidth]{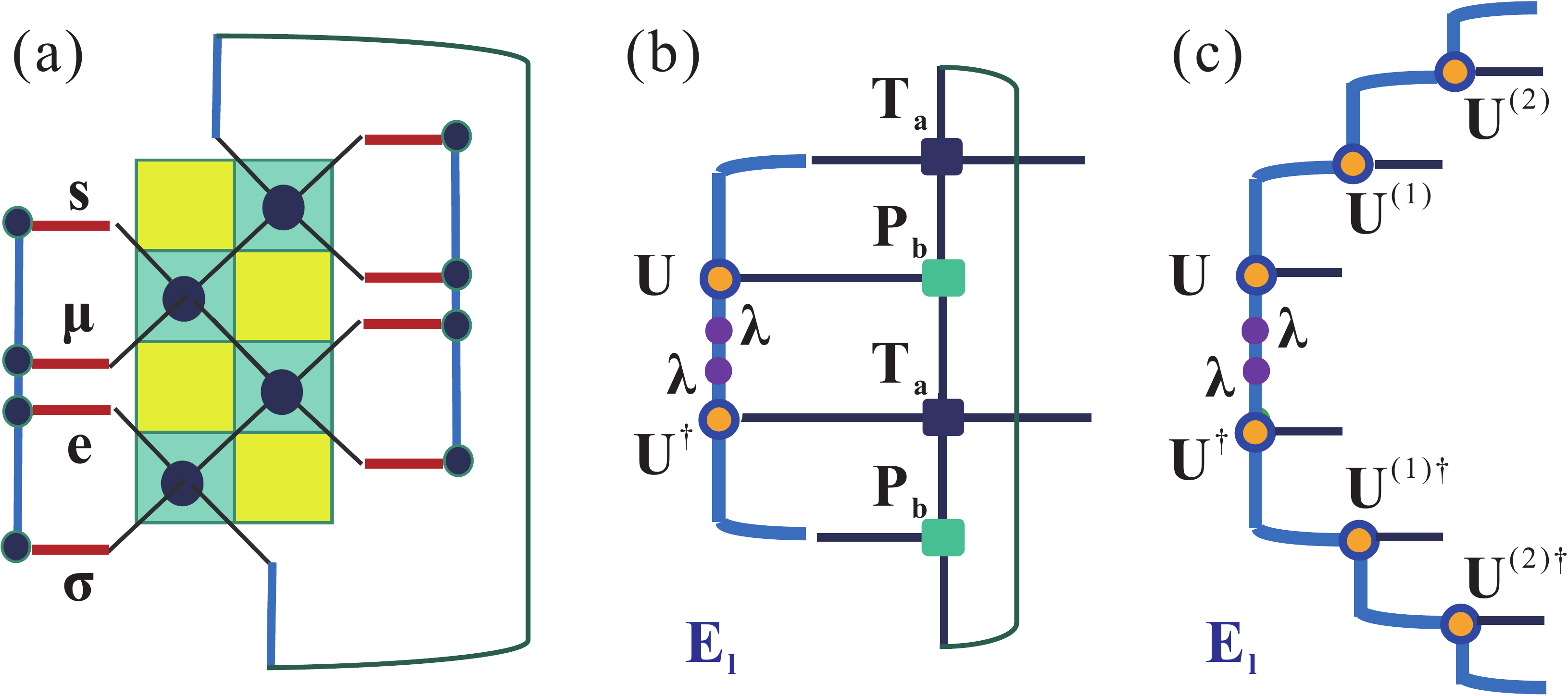}
  \caption{(a) Traditional TMRG algorithm, four blocks $s, \mu, e, \sigma$ constitute a symmetric construction. (b) Bilayer LTRG++ algorithm for contracting thermal tensor network, $E_l \simeq U \lambda^2 U^{\dagger}$ is the dominant left eigenvector of the transfer matrix, where $\lambda$ is a diagonal matrix obtained in LTRG++. (c) Hidden matrix product state representation of $E_l$ in the LTRG++, revealed explicitly by expanding $E_l$ with $\{U^{(i)}\}$ series [and also $(U^{(i)})^{\dagger}$].}
  \label{Fig:equivalence}
\end{figure}

\subsection{LTRG++ algorithm and interacting fermions}
In this subsection, we discuss how to include the fermionic sign in the TTN to solve the electron Hubbard chains, and also provide some details in implementing abelian/non-abelian symmetries in the algorithm. In LTRG++, we contract the thermal TN into two matrix product operators (MPOs), i.e., the upper and the lower layers. The specific scheme could be very flexible, the one shown in Fig. \ref{Fig:LTRG++}(a) exploits a symmetric construction of upper and lower layers, and can save half of the computational costs during projections.

In order to express the partition function of 1D electronic systems in a TTN form, we start with local interaction terms and construct its exponent tensor $\nu$. In Fig. \ref{Fig:LTRG++}(c) we depict the local hopping term as an example, i.e., $(h_{\rm{hop}})_{i,j} = \Psi_i \cdot \Psi_j^{\dagger }$ [note that the complete hopping term should be $(h_{\rm{hop}})_{i,j} + h.c$], where
\begin{equation}
\Psi = (c_{\uparrow}, c_{-\downarrow})
\end{equation}
is a $q=1/2$ irreducible representation of nonabelian $SU(2)_{\rm{spin}}$ symmetry. This can be seen by calculating $[\hat{S}_z, \hat{c}_{\downarrow}] = \frac{1}{2} \hat{c}_{\downarrow}$ and $[\hat{S}_z, \hat{c}_{\uparrow}] = -\frac{1}{2} \hat{c}_{\uparrow}$, the minus sign before $\hat{c}_{\uparrow}$ comes from the fact that $[\hat{S}^+, -\hat{c}_{\uparrow}] = \hat{c}_{\downarrow}$.\cite{Weichselbaum-2012} Owing to the implementation of symmetries, all links in the Fig. \ref{Fig:LTRG++} is assigned with arrows, labeling the directions of fusion. In particular, the horizontal line in Fig. \ref{Fig:LTRG++}(b) points from left to right, marking the hopping direction of electrons. Interestingly, this horizontal line can also be regarded as the Jordan-Wigner string,\cite{JW} which carries a unit of charge and $S=1/2$ spin.

Therefore, considering the fermionic statistics, the $G$ matrix at the cross point of Jordan-Wigner string with physical index $\sigma_2$ should be: $G=-1$ if $\sigma_2$ is of odd parity (i.e., representing states with odd number of electrons), and $G=1$ otherwise. This is because that the Jordan-Wigner string and physical index $\sigma_2$ swap at the intersection point, if they both carry odd parity one should add a $-1$ sign before the wavefunction. Overall, these irreducible operators, symmetric local tensors, parity operators and fusion arrows can be dealt within the framework of QSpace very conveniently .\cite{Weichselbaum-2012}

Similar to the process we have discussed to calculate the partition function before, we can easily obtain the free energy per site, which reads
\begin{equation}
f = \frac{1}{4n} [\sum_{i=1}^{n} (\log \kappa_i^a + \log \kappa_i^b) + \log(\theta_{\rm{max}})].
\end{equation}
where $\kappa_i^{a(b)}$ (from $i$-th odd and even substeps, respectively)are the renormalization factors, and $\theta_{\rm{max}}$ is the dominant eigenvalue of the transfer matrix.

\subsection{Equivalence of LTRG++ to TMRG}
At this point, it is very interesting to compare the LTRG++ algorithm with the traditional TMRG. As shown in Fig. \ref{Fig:equivalence}(a), TMRG exploits the ``s-$\mu$-e-$\sigma$" construction, truncates the enlarged system (environment) block $s$-$\mu$ ($e$-$\sigma$) with the aid of dominant eigenvectors of transfer matrix, and adds two blocks per iteration. Note that the transfer matrix is non-Hermitian and has two sets (left and right) of eigenvectors. Take the dominant left eigenvector as an example, one can construct the reduced density matrix of $s$-$\mu$ (and that of $e$-$\sigma$), and retain the largest $m$ states (White's rule).

At first glance, TMRG is quite different from the bilayer LTRG++ algorithm proposed here. The former was implemented within the framework of traditional DMRG \cite{TMRG}, while the latter is developed in the language of tensor networks. More importantly, the TMRG is firstly contracted along the spatial direction (and then along the Trotter direction), while in LTRG++, the contraction is performed in the reversed order (first Trotter and then spatial directions). However, a careful analysis reveals that, in fact, these two algorithms are essentially equivalent, despite some minor difference in technical details.

The key observation leading to this conclusion is that there indeed exists a hidden matrix product state (MPS) in LTRG++ algorithm. In Figs. \ref{Fig:equivalence}(b,c), we show explicitly the hidden MPS, which is nothing but the dominant eigenvector of vertical transfer matrix pursued in TMRG calculations. The concrete definition has been shown in Fig. \ref{Fig:equivalence}(b), where we take the left eigenvector $E_l$ as an example. It is shown that $E_l \simeq U \lambda^2 U^{\dagger}$ where $\lambda$ is a diagonal matrix storing the Schmidt coefficients, obtained from singular value decomposition on the specific bond. \cite{LTRG} Given that, we can further expand $E_l$ in terms of the series of $U^{(i)}$ and $(U^{(i)})^{\dagger}$, the truncation matrices generated in each projection step (not stored during the process of computations, though), and arrive at Fig. \ref{Fig:equivalence}(c) which reveals explicitly the hidden MPS representation of dominant eigenvector of transfer matrix.

Therefore, although it seems that LTRG++ and TMRG are following reversed order in contracting the TTN (Trotter or spatial direction first), a more careful analysis, though, reveals that the truncation matrices in LTRG++ constitute a hidden MPS and these two methods are essentially equivalent. This can be attributed to the fact that a full contraction of TTN must consider both directions in equal footing while the order of contractions only matters superficially.

\section{Extended Hubbard Chain Model and Phase Separation}
\label{Sec:Hubbard}
\begin{figure}[tbp]
  \includegraphics[angle=0,width=0.9\linewidth]{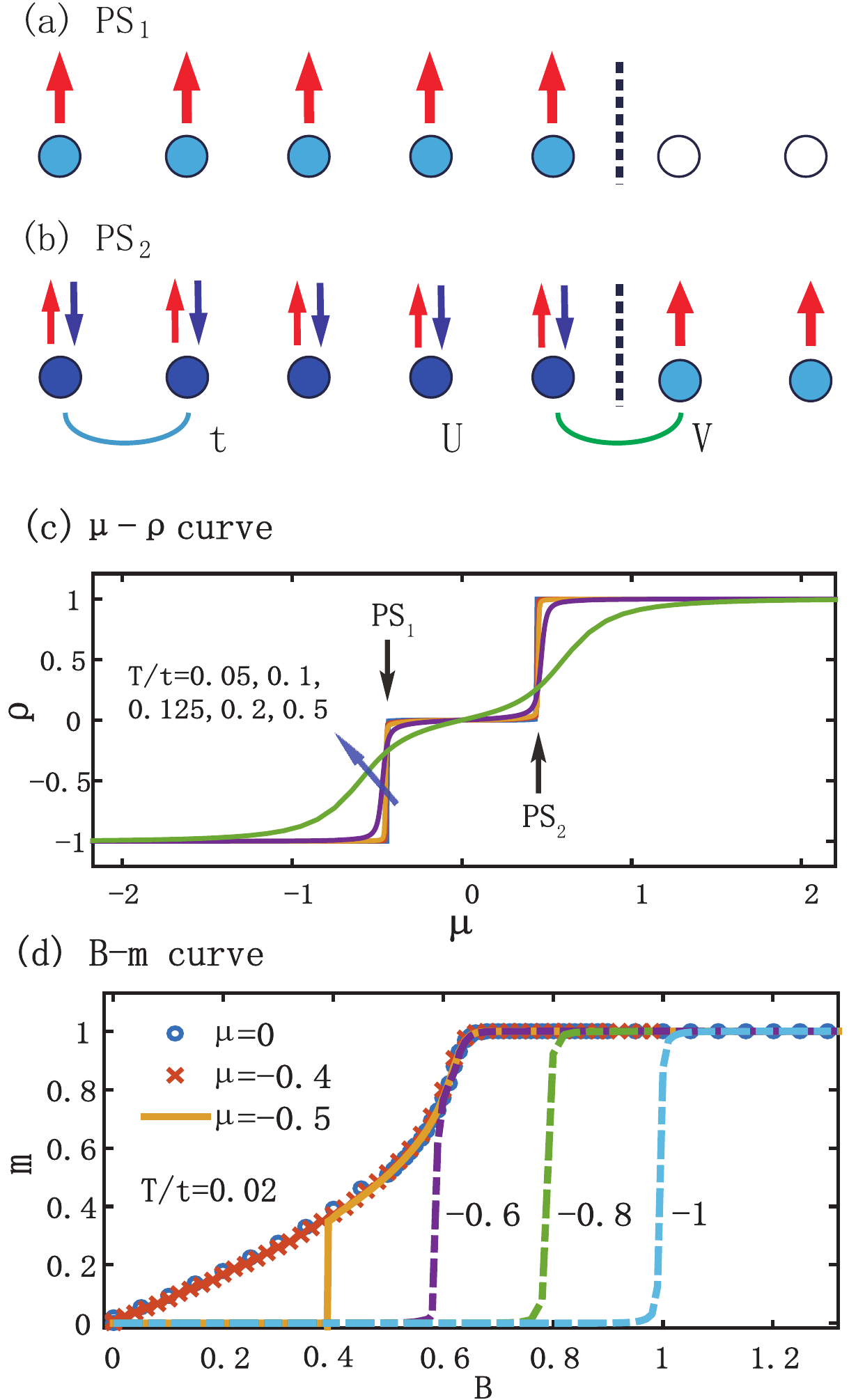}
  \caption{(Color online) Hubbard chain and its groundstate phase separation: (a) between empty (hole) band and a half-filled band (PS$_1$), and (b) between empty (hole) band and doubly occupied band (electron solid) (PS$_2$). $t,U,V$ means nearest-neighbor hopping, on-site and iter-site interaction parameters. (c) The $\mu$-$\rho$ curve of EHM with $U=4,V=-2$, in the absence of magnetic fields, at various temperatures. As $T\to 0$, $\rho$ shows a step-like behavior, two vertical lines correspond to the phase separations in (a) and (b), respectively. (d) The $B$-$m$ curves for various chemical potentials $\mu\leq0$.}
  \label{Fig-Cartoon-PS}
\end{figure}

In this section, we apply the LTRG++ method to solve a fermionic extended Hubbard model (EHM). The Hamiltonian of the EHM with on-site repulsive and inter-site attractive interactions (sometimes also called $t$-$U$-$V$ model) in an external magnetic field reads
\begin{eqnarray}
H = &&-t \sum_{\langle ij\rangle,\sigma} (c_{i,\sigma}^{\dagger} c_{j,\sigma} + h.c.) + \sum_{\langle ij\rangle} U(n_{i, \uparrow}-\frac{1}{2})(n_{i, \downarrow}-\frac{1}{2}) \notag \\
   && +  \sum_{\langle ij\rangle} V \tilde{n}_i \tilde{n}_j - \sum_i[\mu \tilde{n}_i + B (n_{i, \uparrow} - n_{i, \downarrow})],
\label{eq-Hamiltonian}
\end{eqnarray}
where $t=1$ is the hopping amplitude set as energy scale. $\tilde{n} = n-1$ is the particle number operator (relative to half filling), $U$($V$) is the on-site(inter-site) interaction [see Fig. \ref{Fig-Cartoon-PS} (a) for a illustration], $\mu$ is the chemical potential, and $B$ is the external magnetic field. In the followings, we consider exclusively the attractive extended Hubbard chain, i.e., $V<0$.

\subsection{Groundstate Phase Diagram of the EHM}
\label{Sec:HubbardGround}
The groundstate phase diagram of 1D EHM in the absence of magnetic fields has been intensively investigated. For 1D repulsive Hubbard model ($U>0$), there always exists a charge gap at half-filling \cite{Lieb-1968, Essler-2009} for an extended range of $\mu$. When the inter-site interaction $V$ is present, there exist various quantum phases, including spin density wave, charge density wave, bond-order wave, superconductive phases, etc.\cite{Lin-1997, Nakamura-1999, Jeckelmann-2002,Sandvik-2004,Nishimoto-2007} In particular, when $V$ is attractive (i.e., $V<0$), the ground state may show phase separation (PS) phenomenon, i.e., it becomes macroscopically inhomogeneous and can be divided into two separated parts.

In Fig. \ref{Fig-Cartoon-PS}(c), we show the $\rho$-$\mu$ curve of 1D EHM with $U=4$ and $V=-2$, in the absence of magnetic fields. As the chemical potential $\mu$ changes, the average particle number $\rho = \langle \hat{n}-1 \rangle = -\frac{\partial f}{\partial \mu}$ ($f$ is the free energy per site) shows a staircase-like function in the ground state and the step gradually gets blurred with increasing $T$. To be more specific, $\rho=-1$ (meaning empty band) for $\mu \leq -\mu_c $, $\rho=1$ (full band) for $\mu > \mu_c$, and $\rho=0$ (half filling) for intermediate $\mu \in (-\mu_c, \mu_c)$. For the present choice of parameters ($U=4$, $V=-2$), $\mu_c$ is determined to be $\simeq0.44$. Remarkably, from Fig. \ref{Fig-Cartoon-PS}(c), one can see clearly that in the intermediate phase a nonzero charge gap, suggesting the existence of a Mott insulator in this 1D system consisting of itinerant electrons.

In addition, two vertical segments in the staircase curve correspond to two PS states illustrated in Figs. \ref{Fig-Cartoon-PS}(a) and (b), respectively. This observation agrees with the conclusion in Ref. \onlinecite{Lin-1997}, where the Monte Carlo simulations show PS groud states for arbitrary fillings (say, quarter filling, but except for 0, 1/2, or 1 filling), there exist PS ground states. Therefore, Fig. \ref{Fig-Cartoon-PS}(c) shows that the two PS states correspond to two first-order quantum phase transitions (QPTs).

Next, we switch on external magnetic fields, and consider only its Zeeman effect term, i.e., last term in Eq. (\ref{eq-Hamiltonian}). The magnetization curves for various $\mu$ values are shown in Fig. \ref{Fig-Cartoon-PS}(d) at temperature as low as $T/t=0.01$, which are supposed to be quite close to their behaviors in the ground state. Only $\mu\leq0$ curves are shown in Fig. \ref{Fig-Cartoon-PS}(d), the $\mu>0$ cases with the same absolute value $|\mu|$ show the same magnetization curves due to particle-hole symmetry. From Fig. \ref{Fig-Cartoon-PS}(d), we found that the curves show quite different behaviors for different $\mu$ values: for $|\mu| \leq \mu_{c1}$ ($\mu_{c1} \simeq 0.44$), $m$ gradually increases and saturates at $B_s\simeq0.65$, where a second-order QPT occurs; for $\mu_{c1} < |\mu| < \mu_{c2}$, $m$ shows a magnetization jump and then joins the continuous curve which saturate at $B_s$ ($\mu_{c2} \simeq 0.6$ is a tricritical point); for $|\mu|>\mu_{c2}$, a first-order QPT takes place at $B=-U/2-V-\mu=-\mu$, where $m$ jumps from 0 to 1, and there is no more second-order QPT existing. Note that when $|\mu|>\mu_{c1}$, at transition field $B_t$ there exists PS in the ground state. To be specific, for any intermediate $0<m<m_t$, $m_t$ is the magnetization at $B=B_t$, the ground state is a mixture of an empty band and a half-filled band (partially magnetized critical SDW for $\mu_{c1}<|\mu|< \mu_{c2}$ and fully magnetized band for $|\mu|>\mu_{c2}$). This is the PS1 state indicated by left arrow in Fig. \ref{Fig-Cartoon-PS}(c), and also illustrated (in a cartoon way) in Fig. \ref{Fig-Cartoon-PS}(a). Similarly, the right arrow PS2 in Fig. \ref{Fig-Cartoon-PS}(c) is a PS between a full band and a half-filled one, illustrated in Fig. \ref{Fig-Cartoon-PS}(b).

\begin{figure}[tbp]
  \includegraphics[angle=0,width=1\linewidth]{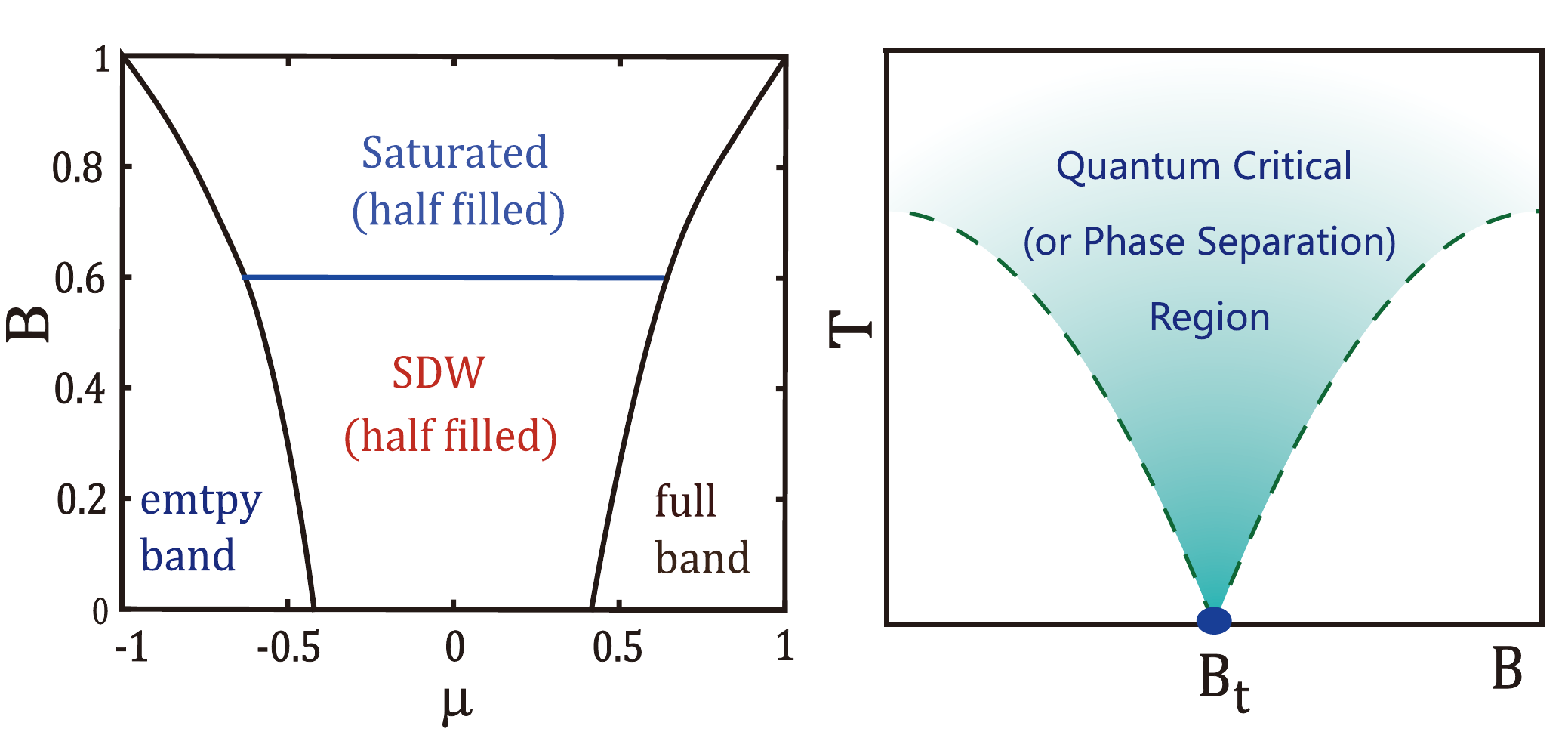}
  \caption{(Color online) (a) Schematic ground state $\mu$-$B$ phase diagram, there exists four different phases at $T=0$. The horizontal (blue) line represent a second-order QPT while the other two (black) curves are phase separation boundaries. (b) Schematic finite-temperature phase diagram of the system, the first- or second-order QPT at transition point $B_t$ has strongly effects on the thermodynamics, for instance MCE properties, in the low-temperature region near it, which is between two dashed lines and highlighted by cyan colors.}
  \label{Fig-PhaseDiag}
\end{figure}

Collecting the transition points, we are able to draw the groundstate phase diagram of 1D EHM under external $B$ fields, shown in in Fig. \ref{Fig-PhaseDiag}(a), which is symmetric along a vertical $\mu=0$ line (due to particle-hole symmetry). There exists four different phases, including hole band, double-occupied (full) band, half-filled critical SDW, and half-filled fully magnetized band. The horizontal blue line is a second-order QPT line, and two curved boundaries representing the PS boundaries. Due to the 1D nature of the model Eq. (\ref{eq-Hamiltonian}), the PS only happens in ground state, i.e., at absolute zero, but it can still have strong influences in low-$T$ properties in the region schematically shown in Fig. \ref{Fig-PhaseDiag}(b). Hereafter, we mainly study the interesting behaviors of MCE properties in this critical or PS region.

\begin{figure}[tbp]
  \includegraphics[angle=0,width=1.0\linewidth]{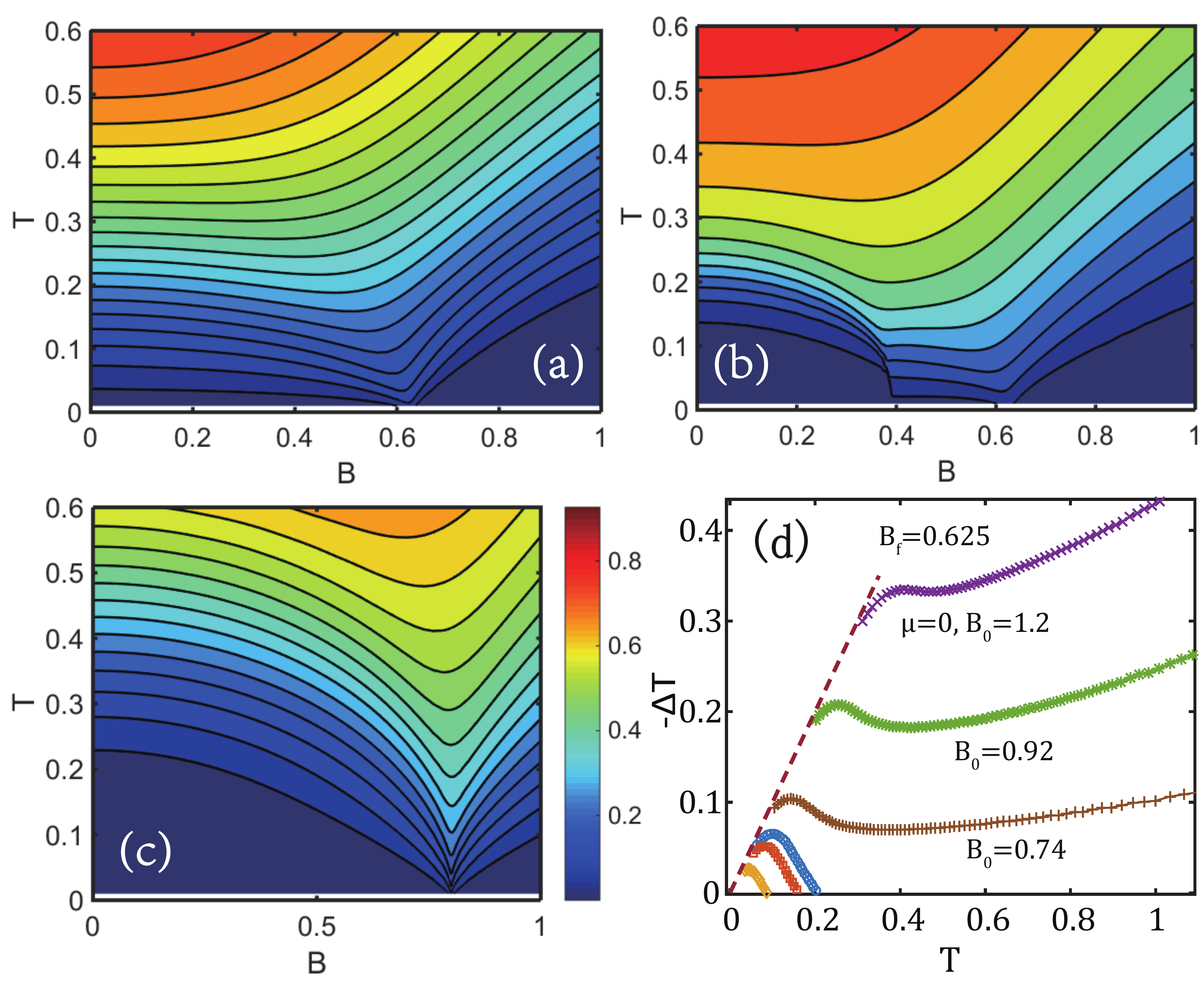}
  \caption{(Color online) (a,b,c) The isentropes of 1D EHM with $\mu=$ 0, -0.5, and -0.8, respectively. Note that they share the same color bar which is shown only in (c). (d) -$\Delta T_{ad} = T(0) - T(B_s)$ extracted from the isentropics is shown versus the initial $T(0)$ for various $\mu$'s. The dashed line shows the asymptotic limit that -$\Delta T_{ad} = T$.}
  \label{Fig-Tad}
\end{figure}

\subsection{Enhanced MCE near the Quantum Critical Point}

In Figs. \ref{Fig-Tad}(a), we show the isentropes in the case of $\mu=0$. The low-entropy curves all show dips around the saturation point $B_s$, where a second-order QPT between SDW and saturated half-filled band takes place. This is a typical example showing that QPT can indeed expands into a quantum critical region [see Fig. \ref{Fig-PhaseDiag}(b)], where the magnetocaloric properties reveal singularity near QPT. To show the adiabatic temperature change $T_{ad}$ more clearly, we follow the isentropic lines in Fig. \ref{Fig-Tad}(a) and plot -$\Delta T_{\rm{ad}} = T(B_0) - T(B=B_f)$ versus initial $T(B_0)$ in Fig. \ref{Fig-Tad}(d), for various $B_0$'s. As initial $T(B_0)$ decreases, we observe that -$\Delta T_{\rm{ad}}$ gradually approaches the the dashed line -$\Delta T = T(B_0)$, i.e., the asymptotic limit that one can never go beyond. However, since our numerical simulation practically always has a lowest temperature $T_{\rm{min}}$ (in this work $T_{\rm{min}}/t=0.01$), our -$\Delta T_{ad}$ curves terminate before it becomes too close to the asymptotic line, i.e., the lower bound is -$\Delta T_{\rm{ad}} \leq T(0)-T_{\rm{min}}$ in the practical simulations. Nevertheless, this asymptotic behavior is guaranteed to be the line isentropes must eventually follow, given that the entropy function $S(T, B)$ is single valued.

In addition, the isentropic contours of 1D EHM with $\mu=-0.5,-0.8$ are also shown in Fig. \ref{Fig-Tad}, from which one can also see a strong influence of first-order QPT on finite-temperature properties, i.e., there also exists a significant change in the formal $T_{\rm{ad}}$ at phase transition points.\cite{Honecker-2005} However, in this case the number of fermions might change in the course of changing magnetic fields, which themselves carry entropies, and the electronic and magnetic entropies can no longer be clearly separated. Therefore, it is tricky to define adiabatic temperature change in this grand canonical ensemble. However, in the case Fig. \ref{Fig-Tad}(a), since $\mu=0$ the total particle number is conserved, and the MCE and adiabatic temperature change therein is well-defined.

\section{Conclusion}
In this paper, we have discussed the linearized tensor renormalization group (LTRG) method, and upgrade it to a bilayer form, which is called LTRG++. This bilayer algorithm, for efficiently contracting the Trotter-Suzuki thermal tensor network, can be connected to several well-known algorithm, making it a very versatile tool and an unified framework to implement these algorithms. For instance, LTRG++ algorithm for finite-size system is a tensor network version of finite-temperature DMRG, and for infinite-size system LTRG++ is essentially equivalent to the TMRG method. We have also discussed the application of LTRG++ to fermionic systems, and employed it to accurately calculate the thermodynamics of a 1D extended Hubbard chain. The enhanced MCE near second-order quantum phase transition points in this fermion chain has also been discovered.

\begin{acknowledgments}
\textit{Acknowledgments.}--- WL is indebted to Andreas Weichselbaum, Tao Xiang, Shi-Ju Ran, Ji-Ze Zhao, Gang Su, Lei Wang and Fei Ye for useful discussions. This work was supported by the National Natural Science Foundation of China (Grant No. 11504014), and the Beijing Key Discipline Foundation of Condensed Matter Physics.
\end{acknowledgments}

\end{document}